\documentclass[aps,showpacs,prl,twocolumn,superscriptaddress]{revtex4}

\usepackage[latin1]{inputenc}  
\usepackage{amsmath}           
\usepackage{amsfonts}         
\usepackage{amssymb}          
\usepackage{graphicx} 
\usepackage{stackrel}
\usepackage{subfigure}

\newcommand{\I}{\mathrm{i}}
\newcommand{\eqRef}[1]{Eq.(\ref{#1})}
\newcommand{\figRef}[1]{Fig.\ref{#1}}	
\newcommand{\tabRef}[1]{Tab.\ref{#1}}	

\newcommand{\CO}[1]{{}}

\DeclareMathOperator{\sgn}{sgn}

\def\XXint#1#2#3{{\setbox0=\hbox{$#1{#2#3}{\int}$ }
\vcenter{\hbox{$#2#3$ }}\kern-.5\wd0}}

\begin{document}

\title{Exact results for the Kondo screening cloud of two helical liquids}

\author{Thore Posske}
\affiliation{Inst. for Theoretical Physics and Astrophysics, Univ. of W{\"u}rzburg, 97074 W{\"u}rzburg, Germany}
\author{Chao-Xing Liu}
\affiliation{Dept. of Physics, The Pennsylvania State Univ., Univ. Park,
Pennsylvania 16802-6300, USA}
\author{Jan Carl Budich}
\affiliation{Inst. for Theoretical Physics and Astrophysics, Univ. of W{\"u}rzburg, 97074 W{\"u}rzburg, Germany} 
\affiliation{Dept. of Physics, Stockholm Univ., Se-106 91 Stockholm, Sweden}
\author{Bj{\"o}rn Trauzettel}
\affiliation{Inst. for Theoretical Physics and Astrophysics, Univ. of W{\"u}rzburg, 97074 W{\"u}rzburg, Germany}



\begin{abstract}
We analyze the screening of a magnetic quantum dot with spin $\frac{1}{2}$ coupled to two helical liquids. Interestingly, we find two qualitatively different sets of Toulouse points, i.e., nontrivial parameters for which we can solve the two channel Kondo model exactly.
This enables us to calculate the temperature and voltage dependent Kondo screening cloud, which develops oscillations for an applied spin voltage $\mu_s$. Such a spin voltage  can be conveniently applied by a charge bias in a four-terminal helical liquid setup.
\end{abstract}

\maketitle

\textit{Introduction.}
The crucial ingredient of Kondo physics \cite{KondoSeminalPaper, HewsonTheKondoProblemToHeavyFermions} is the coupling of a localized spin degree of freedom, often represented by a spin on a quantum dot (QD), to a spin bath. In recent years, peculiarities relating to a plethora of realizations of the spin bath have been investigated \cite{MuellerHartmannZittartzKondoEffectInSuperconductors, 
LeggettEtAlDynamicsOfTheDissipativeTwoStateSystem,
MartinekKondoEffectInQuantumDotsCoupledToFerromagneticLeads, FurusakiNagaosaKondoEffectInATomonagaLuttingerLiquid, MaciejkoLiuOregQiQuZhangKondoEffectInTheHelicalEdgeLiquidOfTheQuantumSpinHallState}.
One of the archetypal phenomena in Kondo physics which is still subject of active research is the characteristic screening of the QD-spin: while theorists predict a macroscopically extended screening which has been coined the Kondo cloud, experimental confirmation of this unique correlation is still lacking \cite{BordaKondoScreeningCloudInAOneDimensionalWire,AffleckSimonDetectingTheKondoScreeningCloudAroundAQuantumDot}. 
Problems related to the direct detection of the Kondo cloud are the high frequency at which the spin of the QD is flipping and the principal inability to directly measure correlations between the QD-spin and the lead-spin without decisively perturbing the tunneling region. 
Helical liquids, which have recently been theoretically predicted \cite{KaneMeleQSHEGraphene, BernevigZhangQSHEInHgTeWells, WuBernevigZhangHelicalLiquidAndTheEdgeOfQuantumSpinHallSystems} and experimentally discovered \cite{KoenigQSHE} at the edge of the quantum spin Hall insulator \cite{KaneMeleQSHEGraphene, BernevigZhangQSHEInHgTeWells, HasanKaneTopInsReview}, feature two modes of excitations with opposite spins moving into opposite directions. 
The quantized conductance of $G = e^2/h$~of a single helical edge state is topologically protected against backscattering by time-reversal symmetry. A magnetic quantum dot coupled to helical liquids is one of the simplest nontrivial perturbations which allows elastic backscattering assisted by a spin flip of the QD \cite{MaciejkoLiuOregQiQuZhangKondoEffectInTheHelicalEdgeLiquidOfTheQuantumSpinHallState}.
Along these lines, former publications \cite{MaciejkoLiuOregQiQuZhangKondoEffectInTheHelicalEdgeLiquidOfTheQuantumSpinHallState, TanakaFurusakiMatveevConductanceOfAHelicalEdgeLiquidCoupledToAMagneticImpurity} have mainly focused on the effect of a magnetic impurity on the conductance of the helical liquid and not on the screening of the localized spin. 
However, attaching helical liquids to a magnetic QD also offers unique opportunities to investigate spin-dependent scattering off the QD: by carrying away spin resolved information
about the Kondo QD in distinct directions, correlations
become measurable away from the QD avoiding the experimental necessity of locally perturbing the tunneling region.

\begin{figure}
\includegraphics[width = .9 \linewidth]{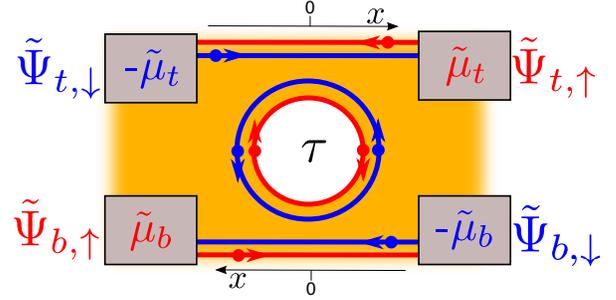}
\caption{(Color online) Two helical liquids of a quantum spin Hall insulator are coupled via an anti-dot. Properly gated, the dot resembles a magnetic impurity with $S = \frac{1}{2}$. Each side is brought out of equilibrium by an applied voltage.}
\label{figSetup}
\end{figure}

In this Letter, we investigate an $S = \frac{1}{2}$ QD that is coupled to two helical liquids in a two channel Kondo model \cite{NozieresBlandinKondoEffectInRealMetals}. We provide exact results for the local screening and the Kondo screening cloud. To this end, we solve the model by restricting the parameters to configurations for which an extension of the method of Emery and Kivelson \cite{EmeryKivelsonKondo} allows us to map the full interacting Hamiltonian to a quadratic one. We call these configurations Toulouse points following the terminology of the noninteracting case. Interestingly, we find two qualitatively distinct sets $A$ and $B$ of Toulouse points; set $A$ contains the Toulouse point of the two channel Kondo model for noninteracting leads \cite{EmeryKivelsonKondo}, while set $B$ resembles a one channel Toulouse point at both Luttinger parameters $g=\frac{1}{2}$ 
\cite{TanakaFurusakiMatveevConductanceOfAHelicalEdgeLiquidCoupledToAMagneticImpurity}.
Determining the local screening and the spatially extended Kondo screening cloud, we are able to demonstrate a different phenomenology for the novel interacting set of Toulouse points. For case $A$, the magnetic field of the QD is always ``perfectly screened``
\cite{ZarandVonDelftAnalyticalCalculationOfTheFiniteSizeCrossoverSprectrumOfTheAnisotropicTwoChannelKondoModel} locally. However, the interacting case $B$ provides, for example, no screening and overscreening.
The Kondo cloud, defined as the correlations between the QD-spin and the lead-spin, obeys an asymptotic decay for large $x$ that is quadratic at zero temperature. At finite temperature $T = 1/(k_B \beta)$, the decay becomes exponential after a length scale $\xi_T \approx \hbar v \beta$, which was also seen by Borda \cite{BordaKondoScreeningCloudInAOneDimensionalWire} for the one channel Kondo model. In contrast to the noninteracting one channel case, we find a $\ln^2(x)$ instead of an $x^{-1}$ divergence for small $x$. 
Furthermore, we observe that an applied spin flavor voltage $\mu_x = \tilde{\mu}_\text{t} + \tilde{\mu}_\text{b}$, whereby ``flavor`` denotes top(t) or bottom(b) (see \figRef{figSetup}), and a spin voltage $\mu_s =\tilde{\mu}_\text{t} - \tilde{\mu}_\text{b}$ have distinct effects on the Kondo cloud.  While the former acts as an artificial magnetic field which decreases the extent of the Kondo cloud significantly, the latter induces spatial oscillations of the Kondo cloud. Hence, the Kondo cloud can be easily manipulated in our four-terminal setup.

\textit{Model.}
A schematic setup is drawn in \figRef{figSetup} where the magnetic QD is realized by a properly gated anti-dot.
The modeling Hamiltonian is the two channel Kondo model for helical liquids and given by the sum of
\begin{align}
&\mathcal{H}_a = \int dx \left(
\ \sum_{\sigma \in \pm} 
v_{F,a} \tilde{\Psi}^\dagger(x)_{a,\sigma}(\sigma \I \partial_x)\tilde{\Psi}(x)_{a,\sigma}  	
\right.
\nonumber
\\
&\left. \ \ \ \ \ \ \ + \frac{g_{4,a}}{2} \tilde{\rho}_{a,\sigma}^2(x) + g_{2,a} \tilde{\rho}_{a,\downarrow}(x) \tilde{\rho}_{a, \uparrow}(x)
\right),
\nonumber
\\
&\mathcal{H}^{\perp}_{K,a} = J^\perp_a \tilde{\Psi}^\dagger_{a,\uparrow}(0) \tilde{\Psi}_{a,\downarrow}(0) \tau^{-} + H.c. 
,
\nonumber
\\
&\mathcal{H}^z_{K,a} =  \frac{1}{2} J_a^z \left(\tilde{\rho}_{a,\uparrow}(0) - \tilde{\rho}_{a,\downarrow}(0) \right) \tau^z
\label{eqTwoChannelHamiltonian}
\end{align}
for each side $a \in \{\text{t},\text{b}\}$.
Here, the operators $\tilde{\Psi}_{a,\sigma}$ and $\tau$  represent the fermions of the leads and  the spin of the QD respectively, $v_{F,a}$ is the Fermi velocity, $g_{2/4,a}$ denotes the interaction strengths within the leads, and the constants $J_a^\sigma$ with $J_a^{x} = J^y_a \equiv J^\perp_a$ determine the strength of the coupling of the leads to the QD.
We mention here that we do not consider ''crossing terms`` of the form $\tau^\lambda \tilde{\Psi}^\dagger_{t,\sigma} \sigma^\lambda_{\sigma,\sigma^\prime} \tilde{\Psi}_{b,\sigma^\prime}$ in our model. This is justified by an RG analysis along the lines of Ref.\cite{LawSengQuantumDotIn2DimTopologicalInsulator}. We show that, for the Toulouse points $B$, all crossing terms are irrelevant and, for the Toulouse points $A$, our model takes into account the most relevant terms (but not all relevant ones); see the appendix for more details.

Following Ref.\cite{VonDelftBosonization}, the helical liquids are bosonized using the bosonization identity
$
\tilde{\Psi}_{a, \sigma}(x) = \frac{1}{\sqrt{2 \pi a_c}}F_{a,\sigma} e^{-\I \tilde{\varphi}_{a,\sigma}(x)}	
$
with the cutoff length $a_c$ and the Klein factors $F_{a,\sigma}$,
and diagonalized by introducing the fields
\begin{align}
\varphi_{a,\pm}(x) =  \frac{1}{\sqrt{8}}
\left[
 \frac{1+g_a}{\sqrt{g_a}} 
  \left( \tilde{\varphi}_{a,\uparrow}(x) \mp \tilde{\varphi}_{a,\downarrow}(-x) \right)
\right.
\nonumber 
\\
\left.  
+
  \frac{1-g_a}{\sqrt{g_a}} 
  \left( \pm \tilde{\varphi}_{a,\uparrow}(-x) - \tilde{\varphi}_{a,\downarrow}(x) \right)
\right]
\label{eqPMFields}	
\end{align}
with the Luttinger parameters
$g_a = \sqrt{v_{J,a}/v_{N,a}}$
and
$v_a = \sqrt{v_{J,a} v_{N,a}}$, whereat $v_{J,a} = v_{F,a} + \frac{g_{4,a}-g_{2,a}}{2\pi}$ and $v_{N,a} = v_{F,a} + \frac{g_{4,a}+g_{2,a}}{2\pi}$.
This leads to
\begin{align}
&\mathcal{H}_a = \int dx \  \frac{v_a}{2} \sum_\sigma \left( \partial_x \varphi_{a,\sigma}(x) \right)^2 
,
\nonumber	
\\
&\mathcal{H}^{\perp}_{K,a} = \frac{1}{4 \pi a_c} J_a^\perp F^\dagger_{a \uparrow} F_{a \downarrow} 
e^{\left( \I \sqrt{2 g_a} \varphi_{a,+}(0)\right)} \tau^- + H.c.
,
\nonumber
\\
&\mathcal{H}^z_{K,a} = 
\frac{J_a^z}{4 \pi}  \sqrt{\frac{2}{g_a}} \partial_x \varphi_{a,+}(0) \tau^z.
\label{eqTransformedHamiltonian}
\end{align}
We furthermore include the chemical potentials $\tilde{\mu}_a$ similarly
to Refs.\cite{SchillerHershfieldKondo,TanakaFurusakiMatveevConductanceOfAHelicalEdgeLiquidCoupledToAMagneticImpurity} by the nonequilibrium operator 
\begin{align}
\mathcal{Y} = \sum_{a \in \{\text{t},\text{b}\} } \tilde{\mu}_a \left(\tilde{\mathcal{N}}_{a,\uparrow} - \tilde{\mathcal{N}}_{a,\downarrow}  \right),
\end{align}
where $\tilde{\mathcal{N}}_{a,\sigma}$ is the total number operator of the fermion $\tilde{\Psi}_{a,\sigma}$. 
In principal, independent chemical potentials at each terminal depicted in \figRef{figSetup} could be considered. However, as the Kondo interaction is a spin interaction, the two potential configurations that are only able to alter charge decouple from the QD. Therefore, we do not consider them explicitly.

\textit{Method.}
The aim is to find the Toulouse points of the Hamiltonian, i.e. points in the parameter space 
for which there is a mapping of the Hamiltonian from \eqRef{eqTransformedHamiltonian} to a quadratic one following the ideas of Refs.\cite{Toulouse1969ExactExpressionOfEnergyOfKondoHamiltonianBaseStateForAParticularJZValue,EmeryKivelsonKondo,ZarandVonDelftAnalyticalCalculationOfTheFiniteSizeCrossoverSprectrumOfTheAnisotropicTwoChannelKondoModel}:
(i)	applying an Emery-Kivelson rotation $U = \exp(\I \sum_{a} \lambda_{a,+} \varphi_{a,+}(0) \tau^z)$,
(ii) transforming $\varphi_{a,\pm}$ orthonormally to fields $\phi_j$,
(iii) refermionizing.
This technique has also been recently employed to analyze the Kondo problem for a single helical liquid in Ref.\cite{Maciejko2012KondoLatticeOnTheEdgeOfATwoDimensionalTopologicalInsulator}.

In our setup, we obtain all Toulouse points by restricting $J^z_a$ 
to the value such that $\mathcal{H}^z_{K,a}$ cancels the Emery-Kivelson rotation of $\mathcal{H}_a$ and secondly imposing that all vertex operators after step (ii) take a refermionizable form.
Here we distinguish two possibilities.
Within case $A$ both vertex operators take the form 
$e^{\pm \I \phi_{4}(0)}$,
while for case $B$ the form of one vertex operator deviates to $e^{\pm \I \phi_{2}(0)}$, where $\phi_{2}$ and $\phi_{4}$ are linearly independent.
A representation of all Toulouse points 
is given in \tabRef{tableToulouseSurfaces}.
\begin{table}
\begin{tabular}{c|c|c|}
&$A$	&$B$
\\
\hline
$v_{\text{t}}, v_{\text{b}}$	& \multicolumn{2}{c|}{$v$}
\\
\hline
$g_{\text{t}}$	&$2 \sin^2(q)$	&$\sin^2(q)$
\\
\hline
$g_{\text{b}}$	&$2 \cos^2(q)$	&$\cos^2(q)$
\\
\hline
$g_{\text{t}}+g_{\text{b}}$	&$2$	&$1$
\\
\hline
$J^z_{\text{t}}$	&$2 \pi v g_{\text{t}}$	&$\pi v(1 - \cos(2 q) +s \sin{(2 q)})$
\\
\hline
$J^z_{\text{b}}$	&$2 \pi v g_{\text{b}}$	&$\pi v(1 + \cos(2 q) +s \sin{(2 q)})$
\\
\hline
\end{tabular}
\caption{Toulouse points of the two channel Kondo model for helical liquids. There are two disconnected sets $A$ and $B$ of Toulouse points whereby $B$ possesses two branches, distinguished by $s = \pm$. The parametrization uses $v \in (0,\infty)$ and $q \in (0,\pi/2)$.}
\label{tableToulouseSurfaces}
\end{table}
Case $A$ is characterized by $g_{\text{t}} + g_{\text{b}} = 2$ and contains the noninteracting case. In contrast, the novel case $B$ obeys $g_{\text{t}}+g_{\text{b}} =1$ and intrinsically relies on interactions. In particular, it contains $g_\text{t} = g_{\text{b}} = \frac{1}{2}$  for which refermionization is known to be a promising method in similar models \cite{VonDelftBosonization, TanakaFurusakiMatveevConductanceOfAHelicalEdgeLiquidCoupledToAMagneticImpurity} and possesses two solvable values of $J^z_{\text{t/b}}$ for each solvable configuration of $g_\text{t/b}$.

The grand canonical operators of the resulting resonant level models after refermionization to the fermionic fields $\Psi_j(x) = \frac{1}{\sqrt{2 \pi a_c}} G_j e^{-\mathrm{i} \phi_j(x)}$ are \cite{EmeryKivelsonKondo}
\begin{align}
\mathcal{H}_A &=
- \mu_x \mathcal{N}_2 - \mu_s \mathcal{N}_4  + \mathcal{H}_0
\nonumber
\\
&+\frac{1}{2 \sqrt{2 \pi a_c}}
\left(
J^\perp_L \Psi^\dagger_4(0) c
+ J^\perp_R  \Psi_4(0) c
+ H.c.
\right)
\label{eqHamiltonianCaseA}
\end{align}
with the local pseudofermion $c = G^\dagger_2 \tau^-$,
for case $A$, where we exploited the extended treatment of Klein factors and number operators in Ref.\cite{ZarandVonDelftAnalyticalCalculationOfTheFiniteSizeCrossoverSprectrumOfTheAnisotropicTwoChannelKondoModel} and  
\begin{align}
\mathcal{H}_B &=
-(\mu_x - \mu_s)\mathcal{N}_2 - (\mu_x + \mu_{s}) \mathcal{N}_4  + \mathcal{H}_0
\nonumber
\\
&+
\frac{1}{2 \sqrt{2 \pi a_c}}
\left(
J^\perp_L \Psi^\dagger_4(0) c
+ J^\perp_R  \Psi^\dagger_2(0) c
+ H.c.
\right)	
\label{eqHamiltonianCaseB}
\end{align}
for case $B$ with $c = \tau^-$, $\mathcal{N}_{2/4} = \frac{1}{2}(\tilde{\mathcal{N}}_{t/b,\uparrow}-\tilde{\mathcal{N}}_{t/b,\downarrow})$ and
$G_{2/4} = F^\dagger_{t/b,\downarrow} F_{t/b,\uparrow}$
similarly to Ref.\cite{VonDelftBosonization}.
In both cases, the noninteracting Hamiltonian is given by
$\mathcal{H}_0 = v \sum_{j}  \int dx \ \Psi_j(x) (\I \partial_x) \Psi_j(x)$.
The Hamiltonians of Eqs.(\ref{eqHamiltonianCaseA}) and (\ref{eqHamiltonianCaseB}) are solvable via a variety of techniques, see, e.g. Refs.\cite{SchillerHershfieldKondo, ZarandVonDelftAnalyticalCalculationOfTheFiniteSizeCrossoverSprectrumOfTheAnisotropicTwoChannelKondoModel}.
Our results are based on infinite order perturbation theory in the Keldysh formalism.

\textit{Results.}
We focus now on the $z$-screening of the spin $\tau$ of the QD. The contributions to this quantity are the local screening at $x=0$, determined by the locally bound spin in the leads, and the spatially extended Kondo cloud.
The local screening $\textstyle S^z_0 = \sum_{a} \int_{-\epsilon}^{\epsilon} dx \ \langle \tilde{\rho}^z_{\text{spin},a}(x) \rangle$, with $\tilde{\rho}_{\text{spin},a} = \frac{1}{2} (\tilde{\rho}_{a,\uparrow}- \tilde{\rho}_{a,\downarrow})$ and $\epsilon = 0^+$, is given by
\begin{align}
S^z_0 &=
- \langle \tau^z \rangle \sum_a \frac{J^z_a}{4 \pi v g_a}.
\end{align}
Following \tabRef{tableToulouseSurfaces}, this reveals $S^z_0 = -\langle \tau^z \rangle$ for case $A$ and the magnetic field of the QD is locally screened.
For case $B$ however, this results in $S^z_0 = - \langle \tau^z \rangle \left(1 + \frac{s}{2 \sqrt{g_{\text{t}} g_{\text{b}}}} \right)$, whereat $s =\pm$ corresponds to the two possible branches of Toulouse points in case $B$.
Hence, for an arbitrary large interaction in one of the helical liquids, the locally accumulated spin gets arbitrarily large as well. This interaction induced phenomenon clearly distinguishes Fermi liquid leads from helical liquid leads.

The Kondo screening cloud $\chi^z_a(x)$ is represented by the spatially resolved correlations between the QD-spin and the lead-spin \cite{MuellerHartmann1969SpinCorrelationInDiluteMagneticAlloys, NagaokaSelfConsistentTreatmentOfKondosEffectInDiluteAlloys}
\begin{align}
\chi^z_a(x,g_{\text{t}},g_{\text{b}}) = \langle  \delta\tilde{\rho}_{\text{spin},a}(x) \delta\tau^z \rangle,
\end{align} 
with $\delta A = A - \langle A \rangle$. 
We find that for case $A$, the Kondo cloud does not depend on the interaction parameters $g_{\text{t/{b}}}$. For convenience, we therefore denote the Kondo cloud of case $A$ by $\chi^z_a(x,1)$ later.
Furthermore, the sum of the Kondo clouds on both sides vanishes, i.e.
\begin{align}
&\sum_a \chi^z_a(x,g_{\text{t}},g_{\text{b}})= 0. 
\label{eqSumRule} 
\end{align}
Interestingly, \eqRef{eqSumRule} also shows that the Kondo correlations of different sides have different signs.
Thereby, the side of positive correlations is determined by the weaker in-plane coupling  $|J^\perp_\text{a}|$. If the in-plane coupling of both sides is equal however, the development of a Kondo cloud is forbidden by this symmetry. 

\begin{figure}
\subfigure
{
\label{CrossoverAndTemperatureKondoCloudA}
\includegraphics[width = .97 \linewidth]{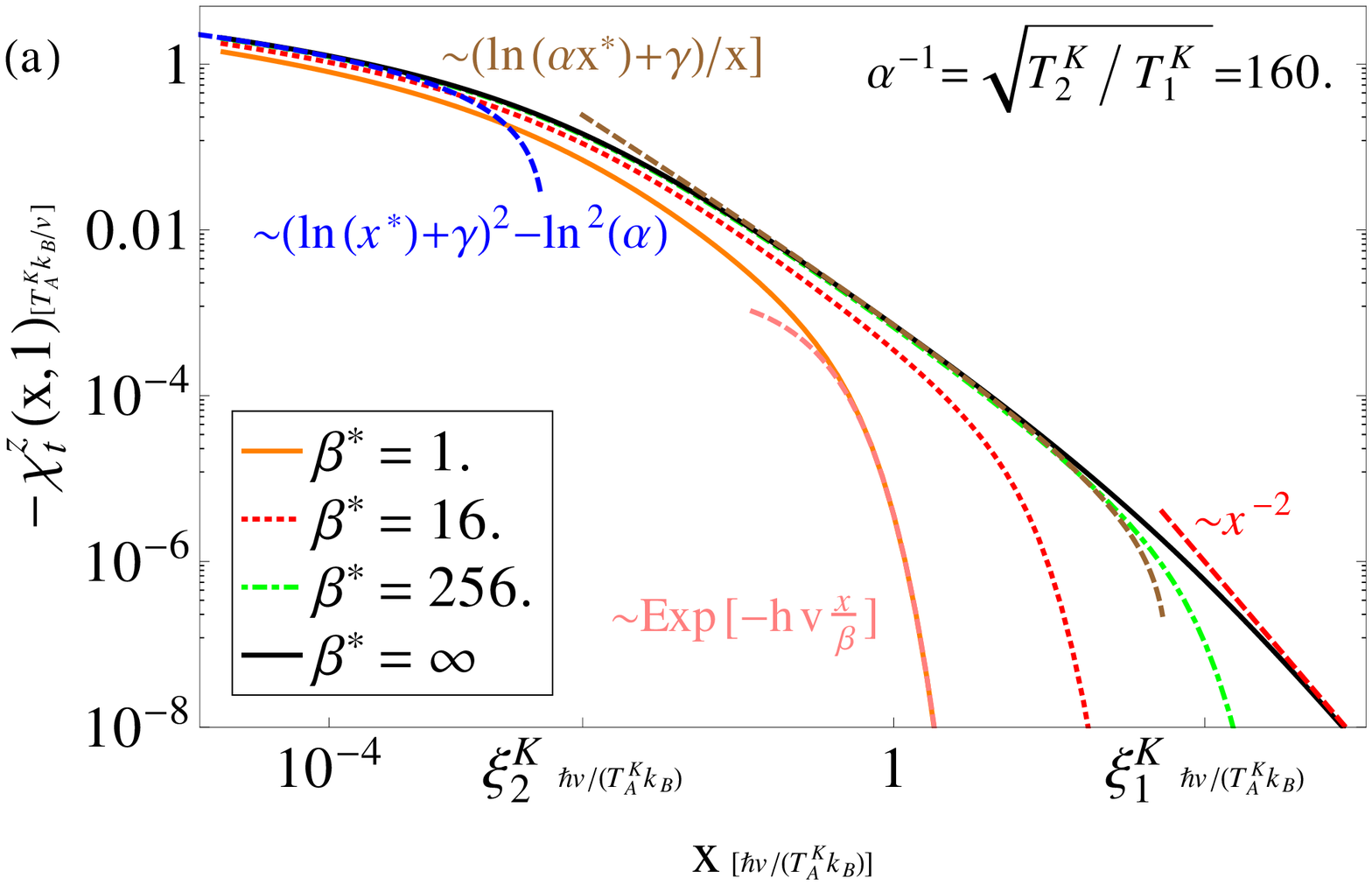}
}

\subfigure
{
\label{CrossoverAndTemperatureKondoCloudB}
\includegraphics[width = .97 \linewidth]{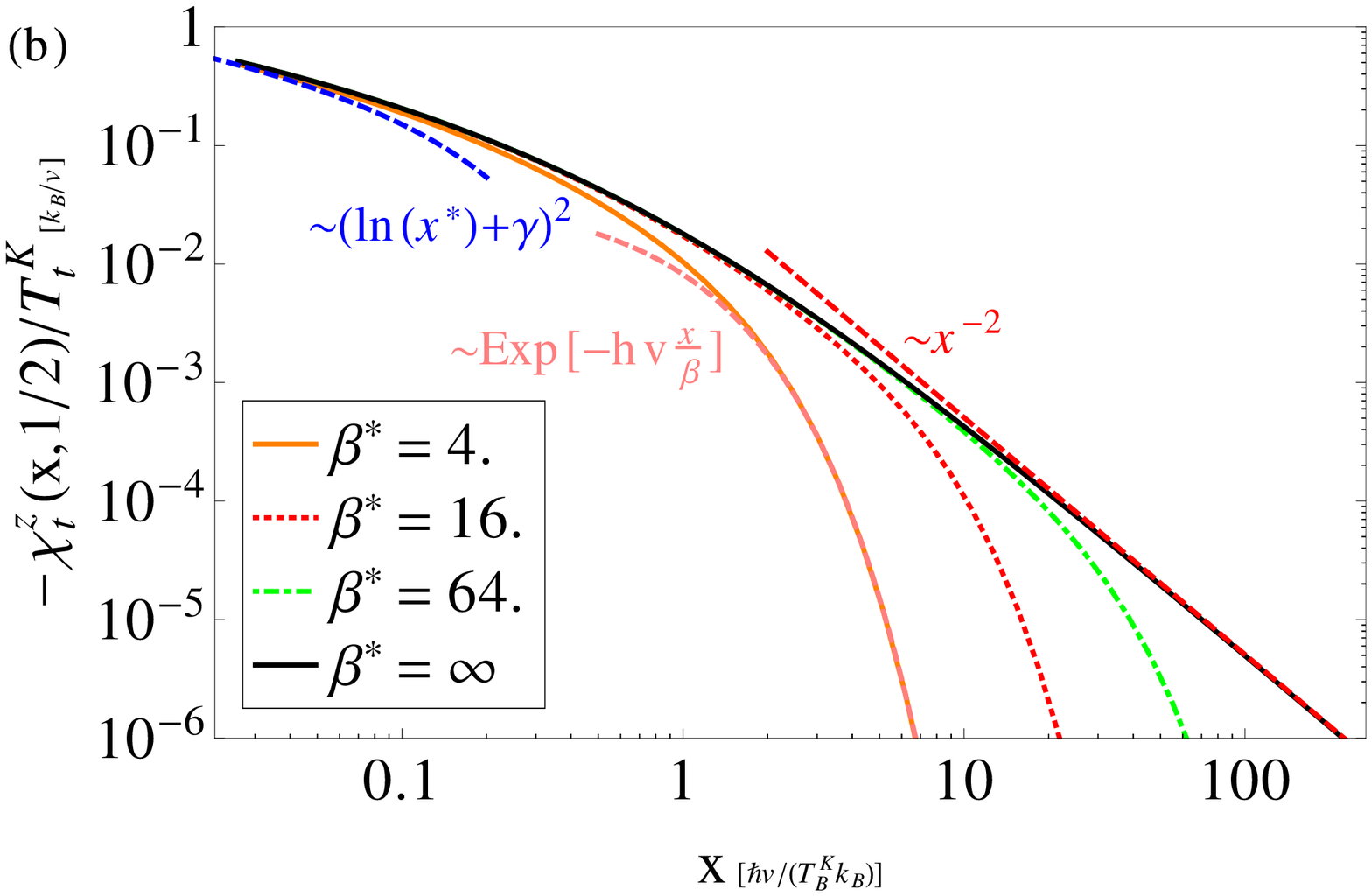}
}
\caption{(Color online) The decay of the generic Kondo clouds in equilibrium for different inverse temperatures $\beta^* = k_B T^K_{A/B} \beta $ on a double-logarithmic scale.
The relevant limits of the cloud are depicted as dashed curves, $\gamma$ is the Euler-Mascheroni constant. 
(a) Noninteracting Kondo cloud. The temperature of reference is $T^K_A$. Both Kondo temperatures $T^K_{1/2}$ determine crossover length scales of the cloud at $\xi^K_{1/2}=\hbar v /(k_B T^K_{1/2})$. 
(b) Interacting Kondo cloud rescaled by $T^K_\text{t}$. The temperature of reference is $T^K_B$. The curve is universal for all $J^\perp_\text{a}$ and only the length scale determined by $T^K_B$ is relevant.}
\label{CrossoverAndTemperatureKondoCloud}	
\end{figure}

For case $B$, all Kondo clouds are derivable from a generic Kondo cloud which is the one of the Toulouse point with $g_\text{t}=g_\text{b}=\textstyle{\frac{1}{2}}$ and $s=-1$. We denote it by $\chi^z_a(x,{\textstyle{\frac{1}{2}}})$.
For arbitrary interaction parameters, the other Kondo clouds are then given by
\begin{align}
\chi^z_a(x,g_{\text{t}},g_{\text{b}}) &=
-\frac{1}{2}
\sum_{a^{\prime}}
\left(
	a a^\prime + s \sqrt{ \frac{g_{-a}}{g_{a}} } 
\right)
\chi^z_{a^\prime}(x,\textstyle{\frac{1}{2}}),
\label{KondoCloudRelationToArchetypCaseB}
\end{align}
with $a,a^\prime \in \{\text{t}\equiv+,\text{b}\equiv-\}$.
A direct consequence of \eqRef{KondoCloudRelationToArchetypCaseB} is a simple dependence of the total Kondo cloud on the interaction parameters
\begin{align}
&\sum_{a} \chi^z_a(x,g_{\text{t}},g_{\text{b}})
=
\frac{-s}{2 \sqrt{g_{\text{t}} g_\text{b}}} 
\sum_a
\chi^z_a(x,{\textstyle{\frac{1}{2}}}).
\end{align}
In equilibrium, the analytical expressions for the Kondo clouds read
\begin{align}
\chi^z_\text{t}(x,1)=
&\sgn{(J^{\perp}_\text{b} - J^{\perp}_\text{t})} \frac{k_B \sqrt{T^K_1 T^K_2}}{4 \pi^2 v} e^{-\frac{2 \pi |x|}{\hbar v \beta}}
\nonumber
\\
&\tilde{\Phi}\left(|x|,\beta,T^K_1\right) \tilde{\Phi} \left(|x|,\beta,T^K_2\right),
\\
\chi_\text{t}^z(x,{\textstyle{\frac{1}{2}}}) &= 
\frac{k_B T^K_\text{t}}{2 \pi^2 v} e^{-\frac{2 \pi |x|}{\hbar v \beta}}
\tilde{\Phi}^2 \left(|x|,\beta,T^K_1+T^K_2\right),
\end{align}
where $\tilde{\Phi}(|x|,\beta,T^K) = \Phi(e^{-\frac{2 \pi |x|}{\hbar v \beta}},1,\frac{1}{2}+\frac{1}{2 \pi} \beta k_B T^K)$ with the Hurwitz-Lerch transcendent $\textstyle \Phi(z,s,\alpha) = \sum_{n=0}^\infty \frac{z^n}{(n+\alpha)^s}$, the Kondo temperatures $T^K_{1/2} = \frac{(J^\perp_{\text{t}} \mp J^\perp_{\text{b}})^2 \hbar}{16 \pi a_c k_B v}$ \cite{SchillerHershfieldKondo} and $T^K_\text{t} =\frac{(J^\perp_{\text{t}})^2 \hbar}{16 \pi a_c k_B v }$.
The equilibrium Kondo cloud is shown in \figRef{CrossoverAndTemperatureKondoCloud}. Concerning case $A$,
given in \figRef{CrossoverAndTemperatureKondoCloudA},
we chose $T^K_A = \sqrt{T^K_1 T^K_2}$ to be the temperature of reference. 
Both Kondo length scales $\xi^K_{1/2} = \hbar v/(k_B T^K_{1/2})$ appear in the shape of the Kondo cloud by inducing a crossover to different asymptotic behaviors. 
The leading divergence for small $x$ is $\ln^2(x)$ and the asymptotic behavior for large $x$ is a quadratic decay.  
For finite temperature, the Kondo cloud decays exponentially in $x$ after a length scale $\xi_T \approx \hbar v \beta$ similar to Ref.\cite{BordaKondoScreeningCloudInAOneDimensionalWire}.

\begin{figure}
\subfigure
{
\label{VoltageKondoCloudA}
\includegraphics[width = .97 \linewidth]{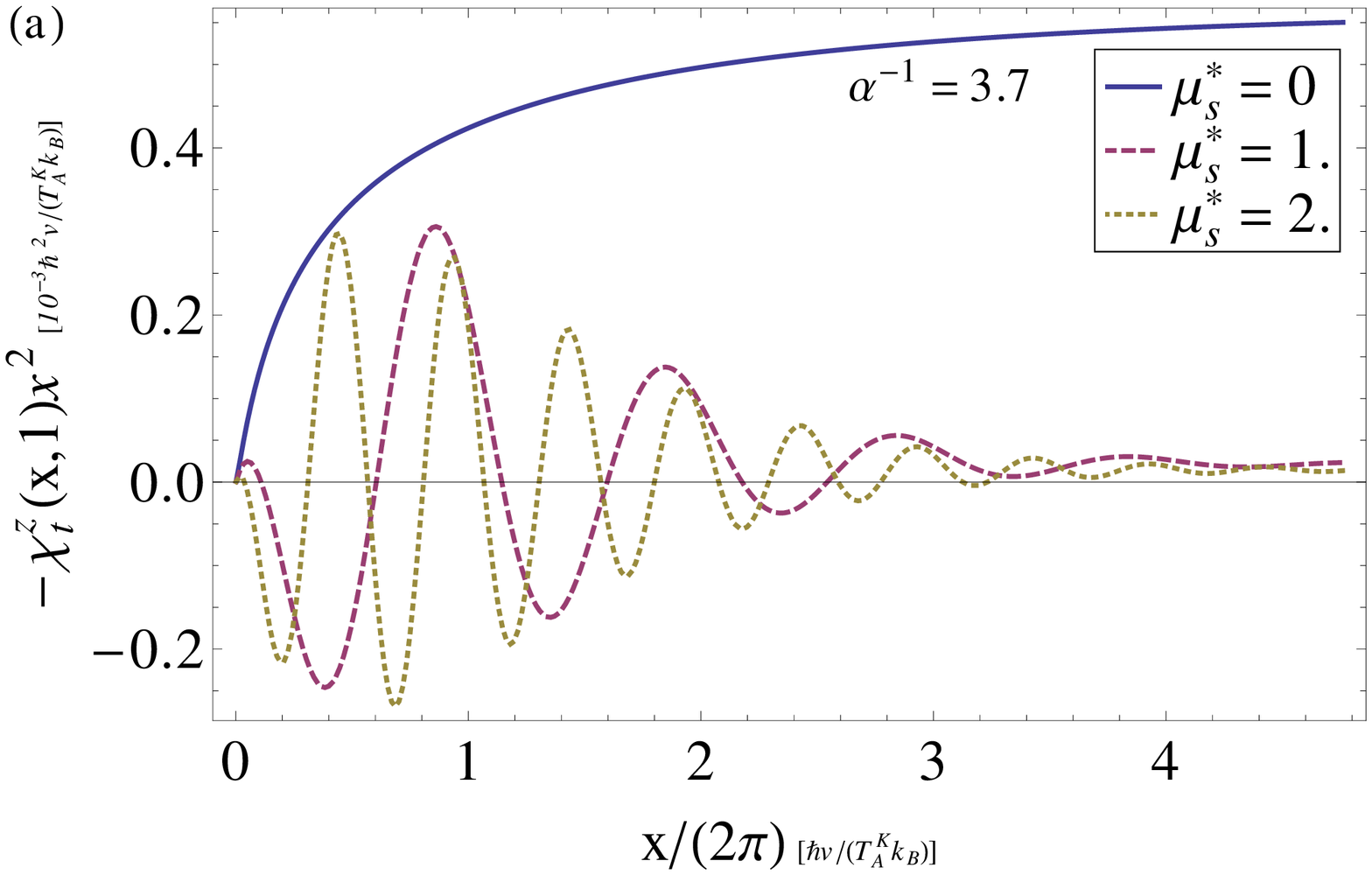}	
}

\subfigure
{
\label{VoltageKondoCloudB}
\includegraphics[width = .97 \linewidth]{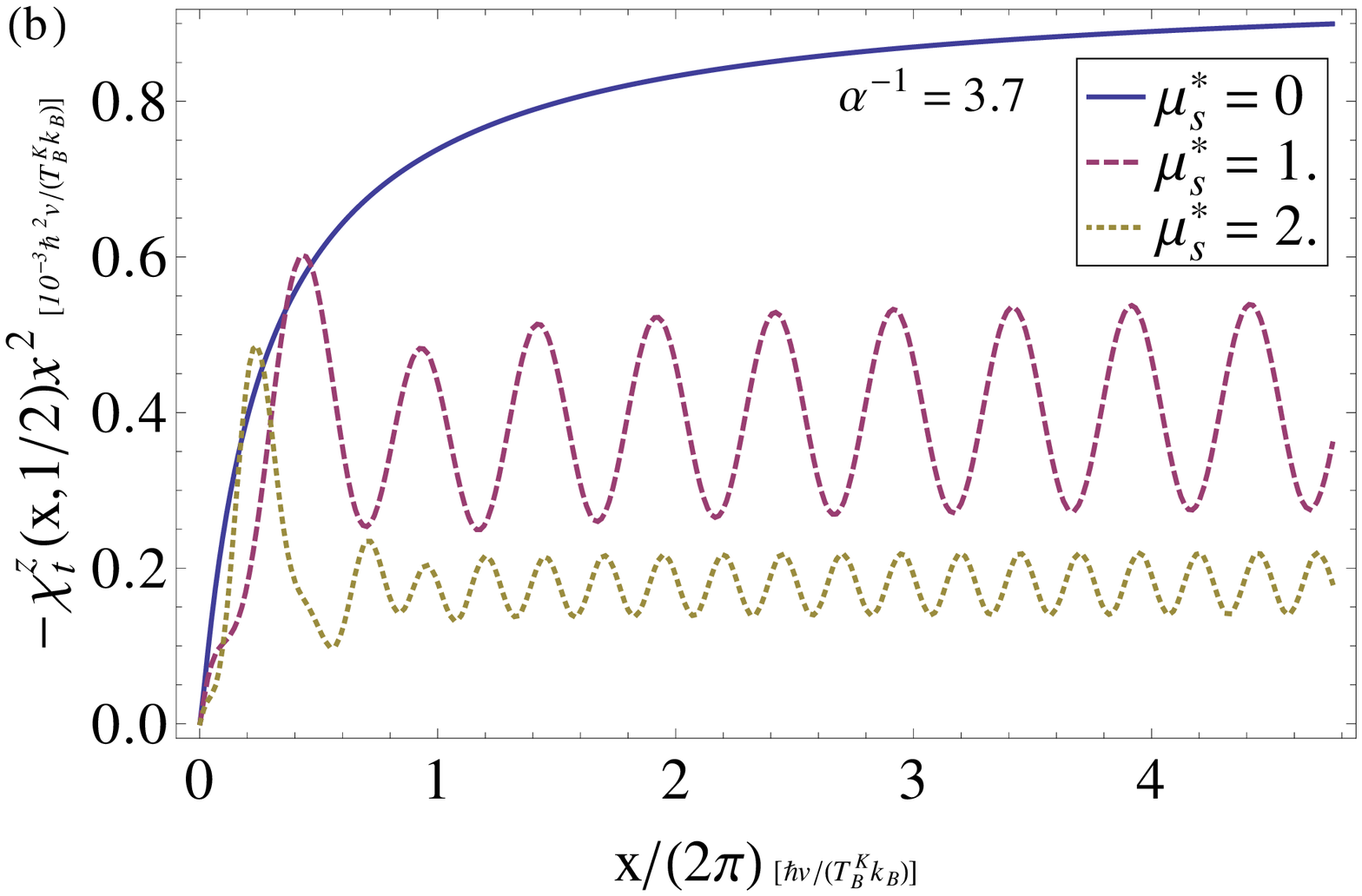}	
}
\caption{(Color online) The generic Kondo clouds rescaled by $x^2$ at zero temperature in nonequilibrium. (a)  Noninteracting Kondo cloud. An applied spin voltage $\mu_s = k_B T^K_{A/B}\mu_s^*$ induces exponentially decaying oscillations of the frequency $\mu_s/(\hbar v)$ in the Kondo cloud and locally changes its sign. (b) Interacting Kondo cloud. Additional permanent oscillations of the frequency $2 \mu_s/(\hbar v)$ appear. 
}
\label{KondoCloudNonEquilibrium}
\end{figure}

For case $B$ (\figRef{CrossoverAndTemperatureKondoCloudB}), the temperature of reference is conveniently chosen to be $T^K_B=T^K_1+T^K_2$. Rescaled by $T^K_\text{t}$, the generic Kondo cloud of side $t$ is a universal curve for all coupling parameters. 
In contrast to case $A$, the only relevant length scale is determined by $\xi^K_B =  \hbar v/(k_B T^K_B)$.

Applied voltages $\mu_x$ and $\mu_s$ alter the Kondo cloud distinctly. The spin flavor voltage $\mu_x$ acts as a magnetic field, see also \cite{TanakaFurusakiMatveevConductanceOfAHelicalEdgeLiquidCoupledToAMagneticImpurity}, and strongly shrinks the Kondo cloud in both classes of Toulouse points.
This reflects the sensitivity of Kondo physics to unhindered spin flips of the QD.
An increasing $\mu_s$ instead induces oscillations in the Kondo cloud depicted in \figRef{KondoCloudNonEquilibrium}. 
For case $A$, an oscillation of the frequency $\mu_s/(\hbar v)$ changes the sign of the Kondo cloud locally. This oscillation decays exponentially and the Kondo cloud becomes monotonic again for large $x$. 
For case $B$, the universality of the Kondo cloud at a specific $T^K_B$ is destroyed. In addition to the oscillations of case $A$, case $B$ shows oscillations of the frequency $2 \mu_s/(\hbar v)$, which decay quadratically in $x$. Nevertheless, the Kondo cloud never changes signs for all $x$.
The oscillations can be interpreted as the impact of Friedel oscillations on the Kondo cloud. 
While Friedel oscillations in helical liquids are suppressed in the presence of ordinary scatterers, the spin-flip scattering off the magnetic QD to the oppositely moving channel generates an interference of the wave functions in both channels.

\textit{Conclusions.}
We couple a magnetic QD with spin $\frac{1}{2}$ to two helical liquids at the edges of a quantum spin Hall insulator.
For the two channel Kondo model, we calculate the Toulouse points at which the Hamiltonian is mapped to a quadratic one.
The Toulouse points separate into two qualitatively different sets. The first is related to the known Toulouse points of the non-interacting case. The second appears exclusively in the case of interactions.
We determine the local screening and the Kondo screening cloud and calculate the signature of applied voltages. 
Due to the special spin-orbit locking of the helical liquids, we conjecture that a finite frequency current noise measurement between different reservoirs (in our four-terminal setup) allows an indirect detection of the Kondo cloud. If the noise frequency matches the Kondo temperature, the current fluctuations should be most sensitive to the Kondo correlations that give rise to the Kondo cloud.

We thank P. Simon and G. Zar{\'a}nd for interesting discussions as well as the DFG, the Humboldt Foundation, and the ESF for financial support.

\appendix

\section{Supplementary material to \textit{``Exact results for the Kondo cloud of two helical liquids''}}
To complement our Letter, we calculate in this supplementary material the scaling dimensions of the crossing terms which are not considered in our model. To this end, we repeat the calculations of Law \textit{et al.} \cite{LawSengQuantumDotIn2DimTopologicalInsulator} in the weak limit of the $xy$-coupling and finite $z$-coupling and extend them to arbitrary interaction strengths on both sides of the quantum spin Hall system. We find that for the majority of Toulouse points the crossing terms become irrelevant, which underlines the experimental relevance of our model.   

Additionally to the in our Letter considered two channel Kondo Hamiltonian, the crossing terms
\begin{align}
\mathcal{H}_{\text{cross}} = \sum_{\lambda \in \{x,y,z\}} J^z \tilde{\Psi}_{\text{t},\sigma}^\dagger(0) \sigma^\lambda_{\sigma,\sigma^{\prime}} \tilde{\Psi}_{\text{b},\sigma^{\prime}}(0) + H.c.
\end{align}
appear in models which assume a one channel contribution. This is, for instance, the case for the effective model of an Anderson impurity  in the Kondo limit \cite{SchriefferWolffTransformation,HewsonTheKondoProblemToHeavyFermions,LawSengQuantumDotIn2DimTopologicalInsulator}. 
Because Anderson models are in many cases the physically correct model for Kondo behavior \cite{NozieresBlandinKondoEffectInRealMetals}, it is an experimentally important question if the crossing terms can be neglected or not. Normally, they can not, which is one of the reasons why two channel Kondo physics is not easily observed.
The situation changes when interactions become important. Imagine a system of two leads attached to an Anderson impurity that is occupied by one particle almost all of the time. If the particle from the impurity tunnels into lead 1, the Coulomb interaction facilitates a subsequent tunneling from lead 1 back to the impurity. Additionally, when a particle from lead 1 tunnels to the impurity, the lower electron density facilitates the subsequent tunneling of a particle back to lead 1. Thereby, the exchange of particles between lead 1 and lead 2 becomes suppressed, which is equivalent to say that the crossing terms become less important.

This concept is proven for many setups. The most important reference for our purposes is Law \textit{et al.} \cite{LawSengQuantumDotIn2DimTopologicalInsulator}, which discusses the scaling dimensions of the crossing terms for our setup but does not consider different interaction strengths for the two helical liquids.
To discuss our case, which crucially relies on the interaction strengths being different, we henceforth extend the calculations of Ref.\cite{LawSengQuantumDotIn2DimTopologicalInsulator} to allow for different interaction strengths. On this basis, we determine the ranges of interaction strengths for which the crossing terms are irrelevant.

Following the notation of our Letter and assuming $J^x_\text{cross} = J^y_\text{cross} \equiv J^\perp_\text{cross}$, the Hamiltonian of the system reads

\begin{widetext}
\begin{align}
\mathcal{H} =& \sum_{a \in {\text{t},\text{b}}}
		\left(
		  \mathcal{H}_{0,a} + \mathcal{H}_{K,a}^\perp + \mathcal{H}_{K,a}^z   
		 \right)
	      + \mathcal{H}_{\text{cross}}^\perp + \mathcal{H}_{\text{cross}}^z,
\\
\mathcal{H}_{0,a} =& \sum_{\sigma} \int dx \frac{v}{2} 
	      (\partial_x \varphi_{a,\sigma}(x))^2,
\\
\mathcal{H}^\perp_{K,a} =&
	      \frac{J^\perp_a}{4 \pi a_c} F^\dagger_{a,\uparrow} F_{a,\downarrow} e^{
		  \left(
		    \I \sqrt{2 g_a} \varphi_{a,+}(0)
		  \right)}
		\tau^{-}
		+ H.c.,
\\
\mathcal{H}_{K,a}^z =&
	      \frac{J^z_a}{4 \pi} \sqrt{\frac{2}{g_a}} \varphi_{a,+}(0) \tau^z,
\\
\mathcal{H}^\perp_{\text{cross}} =&
	      \frac{J^{\perp}_2}{4 \pi a_c} 
	      \left(
	      F^\dagger_{\text{t}, \uparrow} F_{\text{b},\downarrow}
	      e^{ 
		      \frac{\I}{\sqrt{2}}
		      \left(
			\sqrt{g_\text{b}} \varphi_{b,+}(0) + \sqrt{g_\text{t}} \varphi_{t,+}(0) + \frac{1}{\sqrt{g_t}} \varphi_{\text{t},-}(0)-\frac{1}{\sqrt{g_\text{b}}} \varphi_{\text{b},-}(0)
		      \right)} 
\right.
\nonumber
\\
&\left.
	      +
	      F^\dagger_{\text{b}, \uparrow} F_{\text{t},\downarrow}
	      e^{ 
		      \frac{\I}{\sqrt{2}}
		      \left(
			\sqrt{g_\text{b}} \varphi_{b,+}(0) + \sqrt{g_\text{t}} \varphi_{t,+}(0) - \frac{1}{\sqrt{g_t}} \varphi_{\text{t},-}(0) + \frac{1}{\sqrt{g_\text{b}}} \varphi_{\text{b},-}(0)
		      \right)} 
	      \right)
	      \tau^-
	      + H.c.,
\\
\mathcal{H}_{\text{cross}}^z =&
	       \frac{J^z_2}{4 \pi a_c} 
		 \left(
		    F^\dagger_{\text{b},\uparrow} F_{\text{t},\uparrow} e^{
		    \frac{\I}{\sqrt{2}}
		    \left(
		      - \sqrt{g_\text{t}} \varphi_{\text{t},+}(0) - \frac{1}{\sqrt{g_\text{t}}} \varphi_{\text{t},-}(0) + \sqrt{g_\text{b}} \varphi_{\text{b},+}(0) + \frac{1}{\sqrt{g_\text{b}}} \varphi_{\text{b},-}(0)
		    \right)
		    } 
\right.
\nonumber
\\
&\left.
		    +
		    F^\dagger_{\text{b},\downarrow} F_{\text{t},\downarrow} e^{
		    \frac{\I}{\sqrt{2}} 
		    \left(
		      \sqrt{g_\text{t}} \varphi_{\text{t},+}(0) - \frac{1}{\sqrt{g_\text{t}}} \varphi_{\text{t},-}(0) - \sqrt{g_\text{b}} \varphi_{\text{b},+}(0) + \frac{1}{\sqrt{g_\text{b}}} \varphi_{\text{b},-}(0)
		    \right)
		    }
		    \right)
		    + H.c.\ .
\end{align}
\end{widetext}

\begin{table*}[!ht]
\begin{tabular}{c|c|c|c|c|}
Op.{\textbackslash}FPs	&FP general &FP weak	&FP A & FP B \\
\hline
$J^{\perp}_{a}$	&$g_a - \frac{J^z_a}{2\pi v_a} + \frac{1}{16 \pi^2} \sum_{a^\prime} \frac{(J^z_{a^\prime})^2}{g_{a^\prime} \pi^2 v_{a^\prime}^2}$ 	&$\frac{1}{2}(g_{\text{t}}+g_{\text{b}})$	&$\frac{1}{4}(g_{\text{t}}+g_{\text{b}}) =\frac{1}{2}$	&$\frac{1}{2}(g_{\text{t}}+g_{\text{t}}) = \frac{1}{2}$	\\
$J_2^z$	&$\frac{1}{4}(g_{\text{t}}+\frac{1}{g_{\text{t}}}+g_{\text{b}} + \frac{1}{g_{\text{b}}})>1$	&$\frac{1}{4}(g_{\text{t}}+\frac{1}{g_{\text{t}}}+g_{\text{b}} + \frac{1}{g_{\text{b}}})>1$	&$\frac{1}{4}(g_{\text{t}}+\frac{1}{g_{\text{t}}}+g_{\text{b}} + \frac{1}{g_{\text{b}}})>1$	&$\frac{1}{4}(g_{\text{t}}+\frac{1}{g_{\text{t}}}+g_{\text{b}} + \frac{1}{g_{\text{b}}})>1$	\\
$J_2^{\perp}$	&$\frac{1}{4} \sum_{a^\prime}(g_{a^\prime} +\frac{1}{g_{a^\prime}} - \frac{J^z_{a^\prime}}{\pi v_{a^\prime}} + \frac{(J^z_{a^\prime})^2}{4 g_{a^\prime} \pi^2 v_{a^\prime}^2})$	&$\frac{1}{4}(g_{\text{t}}+\frac{1}{g_{\text{t}}}+g_{\text{b}} + \frac{1}{g_{\text{b}}})>1$	&$\frac{1}{4}(\frac{1}{g_{\text{t}}}+\frac{1}{g_{\text{b}}}) >\frac{1}{2}$	 &$\frac{1}{4}(g_{\text{t}}+\frac{1}{g_{\text{t}}}+g_{\text{b}} + \frac{1}{g_{\text{b}}})>1$	\\	
\end{tabular}
\caption{
\label{tabScalingDimensions}
Table of scaling dimensions for our setup following Law \textit{et al.} in Ref.\cite{LawSengQuantumDotIn2DimTopologicalInsulator} and extending their method to different interaction strengths within the leads. FPs stands for ``fixed points''. ``FP general`` is the fixed point for arbitrary interaction strength on both sides, finite $J^z$'s and small $J^\perp$'s. ``FP weak`` is the fixed point of all $J$'s being small. ''FP A'' and ''FP B'' are the special Toulouse points for which the scaling dimensions are calculated by plugging the particular parameters of Tab.I of our Letter into the formulas for the general case.}
\end{table*}
\noindent
To determine whether the crossing terms in $\mathcal{H}^\perp_\text{cross}$ and $\mathcal{H}^z_\text{cross}$ are important we calculate their scaling dimension for a given parameter configuration.
We could use the bare Hamiltonians $\mathcal{H}_{0,a}$ of each side $a$ as the fixed point Hamiltonians and calculate the scaling dimensions for all coupling parameters (other than the $g_\text{t/b}$) being small.
With a bare Hamiltonian of ${\textstyle \sum_a \mathcal{H}_{0,a}}$ we can exploit a useful shortcut to calculate the scaling dimension of vertex operators:
\textit{
With a bare Hamiltonian $\mathcal{H}_0	= \frac{v}{2} \sum_a \int dx (\partial_x \varphi(x))^2$, the scaling dimension $s$ of a vertex operator $V = e^{\sum_a \alpha_a \varphi(0)}$ is $s= \frac{1}{2} \sum_a \alpha_a^2$ in the weak coupling limit.
}
%
With this, we can read off the scaling dimensions for the weak coupling fixed point directly.

However, taking $\sum_a \mathcal{H}_{a,0}$ as fixed point Hamiltonian has the drawback of yielding poor results for the scaling dimensions at some of the $J$'s being finite.
A more accurate approach is to allow for as many finite coupling constants as possible. As indicated in Ref.\cite{EmeryKivelsonKondo}, the model at hand allows an exact solution for all $J^z_{a}$ being finite and all other $J$'s vanishing. 
To this end, a unitary Emery-Kivelson rotation
\begin{align}
\mathcal{H} &\rightarrow U \mathcal{H} U^\dagger = \mathcal{H}^\prime,\\ 
U &= e^{\I \sum_a \lambda_a \varphi_{a,+}(0) \tau^z},\ 
\lambda_a = \frac{J^z_a}{\sqrt{8 g_a} \pi v_a}
\end{align}
is performed.
This absorbs $\mathcal{H}^z_a$ into $\mathcal{H}_{0,a}$ and transforms
\begin{align}
(\tau^{-})^\prime = U \tau^{-} U^\dagger = \tau^{-} e^{- \I \sum_a \lambda_a \varphi_{a,+}(0)}.
\end{align}
All other parts of the Hamiltonian remain unchanged as they commute with $U$ or only give a constant as an additional contribution.
Effectively, the transformation changes the exponents and thereby the first order scaling dimensions of the vertex operators in $\mathcal{H}^\perp_a$
and $\mathcal{H}^\perp_\text{cross}$ while the other operators are unaffected. 
This allows a direct determination of the scaling dimensions with the fixed point Hamiltonian $\sum_a \mathcal{H}_{a,0} + \mathcal{H}^z_{K,a}$.
An overview is given in \tabRef{tabScalingDimensions} where the scaling dimensions of the crossing terms in the general case and at the special Toulouse points are listed. The values for the Toulouse points are obtained by accounting for the special values of $J^z_a$ as described in Tab.I of our Letter.

%

As can be seen, the crossing terms normalize to zero for the weakly coupled impurity and for the Toulouse points of case $B$. For the Toulouse points of case $A$, the situation is more complex. While the $z$-crossing term is irrelevant for the whole parameter range, the $\perp$-crossing term becomes relevant if $1-\sqrt{1/2} \leq g_{\text{t}} \leq 1+\sqrt{1/2}$.
For this regime, we derive from \tabRef{tabScalingDimensions} that the scaling dimension of the $\perp$-crossing term is greater than the scaling dimension of the term that  couples to $J^\perp_a$ (which was also emphasized in Ref.\cite{LawSengQuantumDotIn2DimTopologicalInsulator}). 
This indicates that the coupling of the $\perp$-crossing term grows slower than  $J^\perp_a$ following the RG flow. Therefore, starting at approximately equal strength of both coupling constants, we expect the $J^\perp_a$ term to dominate the crossing term for sufficiently low temperature and applied voltage. Regarding this fact, we still expect our model to be physically meaningful within the range of interactions for which the $\perp$-crossing terms become relevant.

Summarizing, the coupling of the crossing terms are normalized to zero in the following cases:
\begin{itemize}
	\item the weak coupling limit;
	\item all Toulouse points of case $B$;
	\item the Toulouse points of case $A$ for $g_{\text{t}}$ or $g_{\text{b}}$ smaller than $1-\sqrt{1/2}$;
\end{itemize}
and we consider the most relevant terms in our model within the critical parameter region $1-\sqrt{1/2} \leq g_{\text{t}} \leq 1+\sqrt{1/2}$ for the Toulouse points of case $A$.

\bibliography{BibPaperKondoScreeningCloudForHTLLsSubmittedV1p3}

\end{document}